\documentclass[aps,pra,superscriptaddress,showpacs,showkeys,twocolumn,10pt,floatfix]{revtex4-1}

\usepackage{graphicx}
\usepackage{amsmath,amssymb,bm} 
\usepackage{subfigure}

\newcommand{\mv}[1]{\boldsymbol{#1}}
\newcommand{\nn}{\nonumber}

\newcommand{\sref}[1]{Sec.~\ref{#1}}
\newcommand{\eref}[1]{(\ref{#1})}
\newcommand{\fref}[1]{Fig.~\ref{#1}}
\newcommand{\aref}[1]{App.~\ref{#1}}

\DeclareMathAlphabet{\mycal}{OMS}{cmsy}{m}{n}
\DeclareMathOperator{\Tr}{Tr}

\allowdisplaybreaks

\begin{document}

\title{Quantum state transfer through time reversal of an optical channel}

\author{M. R. Hush}
\affiliation{School of Engineering and Information Technology,
University of New South Wales at the Australian Defence Force Academy, Canberra, ACT 2600, Australia}
\email{m.hush@unsw.edu.au}
\author{C. D. B. Bentley}
\affiliation{Department of Quantum Science, Research School of Physics and Engineering, The Australian National University, Canberra, ACT 2601, Australia}
\altaffiliation{Current address: Max-Planck-Institute for the Physics of Complex Systems, D-01187 Dresden, Germany}
\author{R. L. Ahlefeldt}
\affiliation{Department of Physics, Montana State University, Bozeman, MT 59717, USA}
\affiliation{Laser Physics Centre, Research School of Physics and Engineering, The Australian National University, Canberra 2601, Australia}
\author{M. R. James}
\affiliation{Research School of Engineering, Australian National University, Canberra, ACT 2601, Australia}
\affiliation{ARC Centre for Quantum Computation and Communication Technology, Research School of Engineering, Australian National University, Canberra, ACT 2601, Australia}
\author{M. J. Sellars}
\affiliation{Laser Physics Centre, Research School of Physics and Engineering, The Australian National University, Canberra 2601, Australia}
\affiliation{ARC Centre for Quantum Computation and Communication Technology, Research School of Engineering, Australian National University, Canberra, ACT 2601, Australia}
\author{V. Ugrinovskii}
\affiliation{School of Engineering and Information Technology,
University of New South Wales at the Australian Defence Force Academy, Canberra, ACT 2600, Australia}

\begin{abstract}
Rare earth ions have exceptionally long coherence times, making them an excellent candidate for quantum information processing. A key part of this processing is quantum state transfer. We show that perfect state transfer can be achieved by time reversing the intermediate quantum channel, and suggest using a gradient echo memory (GEM) to perform this time reversal. We propose an experiment with rare earth ions to verify these predictions, where an emitter and receiver crystal are connected with an optical channel passed through a GEM. We investigate the affect experimental imperfections and collective dynamics have on the state transfer process. We demonstrate superrandiant effects can enhance coupling into the optical channel and improve the transfer fidelity. We lastly discuss how our results apply to state transfer of entangled states.
\end{abstract}
\pacs{42.50.Ex,42.50.Md,78.47.nd,76.30.Kg}
\keywords{quantum state transfer; rare earth ions; time reversal}


\maketitle

\section{Introduction}

High fidelity quantum state transfer will play an important role in quantum information processing \cite{kimble_quantum_2008,divincenzo_physical_2000}. State transfer using a quantum optical channel, in particular, allows high speed transfer over long distances with little loss \cite{ritter_elementary_2012,droste_optical-frequency_2013}. The challenge is to determine a method to coherently transfer the quantum state of an `emitter' system to a `receiver' system. It has been recognized that perfect state transfer can be achieved between two qubits when the equation of motion of the receiver qubit is a time-reversed version of the emitter  \cite{cirac_quantum_1997}. This principle underpinned the earliest proposal for transferring the state between two qubits with an optical channel, where the coupling between the systems and the channel was engineered such that the carrier photon had a time symmetric wavepacket \cite{cirac_quantum_1997}. However,the time symmetric approach does not obviously scale when transferring the state between multi-level systems or ensembles of qubits. We consider a different approach, which in principle will scale, where we perform time reversal of the quantum channel using a Gradient Echo Memory (GEM) \cite{alexander_photon_2006,hetet_electro-optic_2008,longdell_analytic_2008,hush_analysis_2013} to achieve perfect transfer between two ensembles of qubits. Furthermore, we investigate the collective nature of the ensembles of qubits' coupling to the cavity and the possible advantages this may provide with regard to transfer fidelity. 

We propose an implementation of our protocol with rare-earth ion crystals. Rare-earth ion crystals have exceptionally long coherence times: among hyperfine transitions, $T_2$ coherence times can be as long as hours \cite{zhong_optically_2015}.  Furthermore, work with stoichiometric crystals has shown that it is possible to get very strong interactions between nearby ions \cite{ahlefeldt_precision_2013}, making rare earth ions ideal for quantum information processing. Previous quantum state transfer proposals with rare earth ions have targeted individual ions \cite{mcauslan_strong-coupling_2009}. However, addressing individual rare earth ions is challenging, with few demonstrations \cite{siyushev_coherent_2014,xia_all-optical_2015}; typical experiments involve ensembles of ions \cite{zhong_optically_2015}, and most quantum information processing proposals have targeted ensembles \cite{longdell_demonstration_2004,longdell_experimental_2004,wesenberg_quantum_2004,wesenberg_scalable_2007}. Here we show that working with ensembles rather then individual sites may present an advantage when transferring states where all the ions are in an identical quantum state. The collective phenomenon of superradiance can be used improve the efficiency of coupling into the optical channel. This has previously been exploited in experiments to enter the cavity QED regime with an ensemble of rare earth ions and an optical resonator \cite{probst_anisotropic_2013}. The key to achieving strong coupling between a cavity and an ensemble is having a small inhomogeneous linewidth that is below the linewidth of the cavity \cite{williamson_cavity_2014}; inhomogeneous linewidths smaller than the hyperfine level spacing have recently been achieved in stoichiometric crystals \cite{ahlefeldt_ultra-narrow_2016}. 

After the pioneering work of Cirac \emph{et al.} \cite{cirac_quantum_1997}, there has been further experimental \cite{petrosyan_reversible_2009,sillanpaa_coherent_2007,stute_quantum-state_2013} and theoretical work \cite{mcauslan_coherent_2011,boozer_reversible_2007,stace_mesoscopic_2004} on state transfer between individual qubits mediated by an optical channel. Furthermore, there has been analysis of state transfer between quantum oscillators \cite{zhang_quantum-state_2003,yamamoto_quantum_2016,parkins_quantum_1999}. However, there has been little work on state transfer between ensembles of qubits, and previous analysis has only looked at the perturbative regime where the ensemble can be treated as an oscillator \cite{yin_multiatom_2007,duan_long-distance_2001,matsukevich_quantum_2004}. We consider state transfer between two ensembles of qubits, where the initial state is non-perturbative; our aim is to take an ensemble of separable, identical (up to a rotation in phase) qubits in an arbitrary quantum state and transfer this state onto another ensemble.  

The manuscript is structured as follows: in \sref{sec:gentrans}, we demonstrate that the principle of time reversal can be used to transfer the state between two identical quantum systems, up to a sign change in the Hamiltonian, as long as they have unique dark states. In \sref{sec:GEMtrans}, we demonstrate that a GEM can be used to physically realize time reversal of a coherent pulse in an optical quantum channel. In \sref{sec:reitrans}, we present a proposal for quantum state transfer between two ensembles of rare earth ions, examine the possible advantages of superradiant coupling, and discuss what affects the fidelity of the transfer. In \sref{sec:entangledstates}, we discuss a few ways of extending our approach to transferring entangled states, and where further engineering may be required. Finally, in \sref{sec:discus} we discuss implementation considerations for our transfer scheme and its impact in the context of quantum control. 

\section{Generic quantum state transfer through time reversal} \label{sec:gentrans}

Here we demonstrate that the time-reversal of a quantum channel can be used to perform perfect state transfer between two quantum systems with a unique dark steady state and identical Hamiltonians, up to a sign change. We consider two systems: an `emitter' and a `receiver'. The emitter has a Hamiltonian $H_{\rm em} = H$ and coupling operator $L_{\rm em} = L$. We assume the emitter has a dark pure state, meaning $L|\psi_{ss} \rangle =0$ and $H|\psi_{ss} \rangle = h|\psi_{ss}\rangle$, which is the unique steady state of the system \cite{kraus_preparation_2008}. We can write down the Langevin equation for an arbitrary Hermitian operator of the emitter system: 
\begin{align}
\frac{dX_{\rm em}(t)}{dt} = & -i X_{\rm em} H_{\rm em} \nn \\
& - [X_{\rm em},L_{\rm em}^\dag] ( L_{\rm em}/2 + b_{\rm em,in}(t)) +\mbox{a.t.} \label{eqn:emitter}
\end{align}
where $\mbox{a.t.}$ refers to adjoint transpose of all terms to the left. 
The output of this system will be $b_{\rm em, out}(t) = L_{\rm em}(t) + b_{\rm em, in}(t)$. 
We set the initial state of the system and bath to be $|\Psi_{\rm em} (t=0) \rangle = |\mv{0}\rangle  \otimes |\psi_0\rangle \otimes |\mv{0}\rangle$ where we partition the bath at a time $t$ into an input before the system and an output after the system, and both are initially in a vacuum state $|\mv{0} \rangle$. The system has a unique dark state, so we can be certain it will asymptotically approach a separable state $|\Psi_{\rm em} (t=T) \rangle = |\mv{0}\rangle  \otimes |\psi_{ss} \rangle \otimes |\mv{\phi}\rangle$, where $|\mv{\phi}\rangle$ is some multi-photon output state, for any pure state initial condition. We assume that the system gets very close to this steady state in a time $T$.  We can approximate that $b_{\rm em,out}(t) \approx b_{\rm em,in}(t)$ for $t \geq T$, meaning the output will no longer be correlated with the system. Thus, we can trace out the emitter system without losing purity of the output state.  Next we consider some receiver system which obeys its own Langevin equation, but we use a different time index $\tau$:
\begin{align}
\frac{d{X}_{\rm re}(\tau)}{d\tau} = & -i X_{\rm re} H_{\rm re} \nn \\
&- [X_{\rm re},L_{\rm re}^\dag] (L_{\rm re}/2 + b_{\rm re, in}(\tau)) + \mbox{a.t.} \label{eqnReciever}
\end{align}
Our aim is to engineer this system such that its evolution is a time reversal of the emitter. Specifically, we want the expectation of the error operator $E(X,\tau) = X_{\rm re}(\tau) - X_{\rm em}(t= T - \tau)$, for an arbitrary $X$, to be zero for $0 \le \tau \le T$. We first set the input of the system to be a sign-changed time reversal of the emitter output: $b_{\rm re,in}(\tau) = -b_{\rm em,out}(t=T - \tau) =- L_{\rm em}(t=T - \tau) -b_{\rm em,in}(t=T - \tau)$. We can write down the equation of motion of $E(X,\tau)$ as follows:
\begin{align}
\frac{dE(X,\tau)}{d\tau} = & \frac{dX_{\rm re}(\tau)}{d\tau} + \frac{dX_{\rm em}(t)}{dt} \nn \\
= & -i X_{\rm re}(\tau) H_{\rm re}(\tau) - i X_{\rm em}(t) H_{\rm em}(t) \nn \\
& - [X_{\rm re}(\tau),L_{\rm re}^\dag(\tau)] (L_{\rm re}(\tau) - L_{\rm em}(t))/2 \nn \\ 
& + ([X_{\rm re}(\tau),L_{\rm re}^\dag(\tau)] - [X_{\rm em}(t),L_{\rm em}^\dag(t)])  \nn \\  
&\times(L_{\rm em}(t)/2 + b_{\rm em, in}(t)) + \mbox{a.t.} \label{eqn:errdyn}
\end{align}
where $t = T - \tau$. In order to ensure $E(X,\tau=0) = 0$ we set the initial condition for the receiver to be the final state of the emitter, in terms of the whole state: $|\Psi_{\rm re}(\tau = 0)\rangle = |\mv{\phi}'\rangle \otimes |\psi_{ss} \rangle \otimes |\mv{0}\rangle$ where $|\mv{\phi}' \rangle$ is the time reversed quantum channel and $|\psi_{ss}\rangle$ is the steady state of both the emitter and receiver. Note this initial state was only possible to prepare because the emitter had a unique dark state. Next to ensure $\dot{E}(X,\tau=0) = 0$ we set $H_{\rm re} = -H$ and $L_{\rm re} = L$. It can then be shown the solution $E(X,\tau) = 0$ for $0 \le \tau \le T$ satisfies \eref{eqn:errdyn}, as this implies $X_{\rm re}(\tau) = X_{\rm em}(t = T - \tau)$. The receiver will evolve in a time reversed manner with respect to the emitter, such that final state of the receiver will be the initial state of the emitter: $|\Psi_{\rm re}(T) \rangle = |\mv{0} \rangle \otimes |\psi_0\rangle \otimes |\mv{0}\rangle$. Hence we can perform perfect state transfer between two identical quantum systems, up to a sign change in the Hamiltonian, with unique dark states by simply time reversing the output of the bath and setting the initial condition of the receiver to be the final condition of the emitter.

\section{Time reversal of an optical channel using a GEM} \label{sec:GEMtrans}

\begin{figure*}[t!]
\centering
\includegraphics[width=0.75\linewidth]{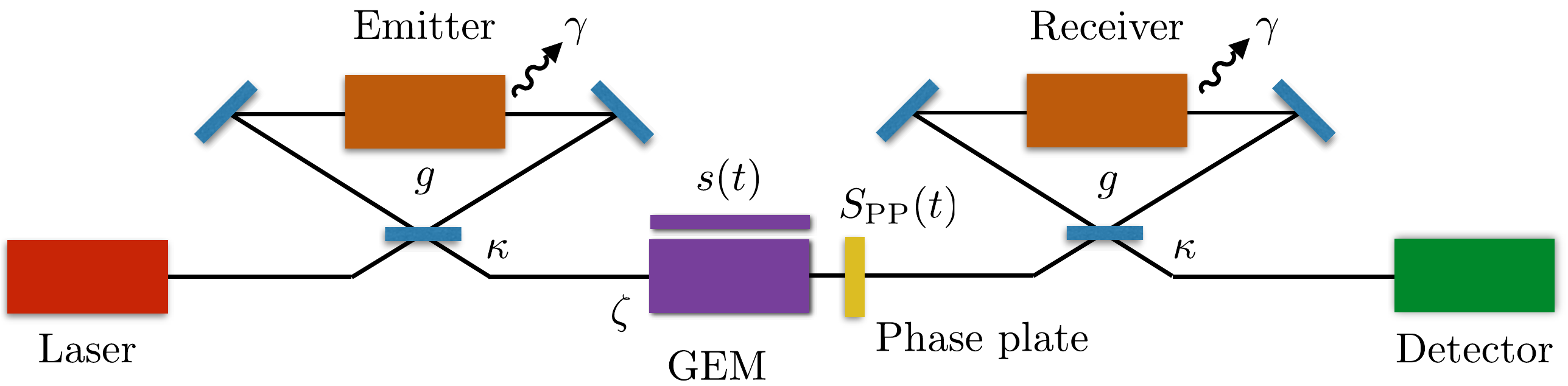}
\caption{An ensemble of rare earth ions termed the `emitter' is coupled, with strength $g$, to a ring cavity, which in turn is coupled, with strength $\kappa$, to a to quantum channel. The quantum channel is passed through a GEM, with a controllable gradient $s(t)$ and optical depth $\zeta$, and phase correction plate $S_{\rm PP}(t)$. The resultant time reversed signal from the emitter is sent to an identical 'receiver' crystal, up to a sign change in the Hamiltonian, where it is perfectly absorbed thus transferring the state. Both the emitter and receiver have a spontaneous emission loss rate of $\gamma$. The laser and detector are used for initial state preparation and measurement of the transfer performance.}
\label{fig:diagram}
\end{figure*}

The next challenge is to determine a physical system that will time reverse a quantum channel. We will use a GEM to achieve this goal. A GEM is described by the following Hamiltonian and coupling operator \cite{hush_analysis_2013}:
\begin{align}
H_{GEM} = & s(t) \int_{-\Xi}^\Xi d\xi \; \xi a^\dag(\xi) a(\xi) \nn \\ 
& + \frac{\zeta}{2i} \int_{-\Xi}^\Xi d\xi \int_{-\Xi}^\xi d\xi' (a^\dag(\xi) a(\xi') - a^\dag(\xi') a(\xi)), \nn \\
L_{GEM} = & \sqrt{\zeta} \int_{-\Xi}^{\Xi} d\xi a(\xi). 
\end{align}
Here $\Xi$ is the bandwidth of the memory, $\zeta$ is the optical depth and $s(t)$ is the sign of the gradient which can be changed between $\pm 1$. 

We can write down the following Heisenberg equations of motion for the atomic excitation operator $a(\xi,t)$ in terms of the input operator $b_{\rm GEM,in}(t)$ and the equation for the output operator $b_{\rm GEM,out}(t)$:
\begin{align}
\dot{a}(\xi,t) = & -is(t) \xi a(\xi,t) + \zeta \int_{-\Xi}^\xi d\xi' a(\xi',t) \nn \\
& + b_{\rm GEM,in}(t), \\
b_{\rm GEM,out}(t) =& \sqrt{\zeta} \int_{-\Xi}^\Xi d\xi' a(\xi,t) + b_{\rm GEM,in}(t). \label{eqn:GEMlang}
\end{align}

The GEM is operated in two stages.  First, a write stage between $t =0$ to $T$, where $s(t) = +1$.  Here the atomic field, initially in a vacuum, stores a light pulse injected into the memory through $b_{in}(t)$. This is followed by the read stage between $t=T$ to $2T$, where $s(t)=-1$.  Here what was stored in the atomic field is read out through $b_{\rm out}$. In the broadband limit, meaning $\Xi$ is much larger than the spectral width of the input pulse, equation \eref{eqn:GEMlang} can be analytically solved as follows \cite{hush_analysis_2013}:
\begin{align}
a(\xi,T) = \chi(\zeta) \int_0^T dt e^{-i \xi t - i \zeta \ln (T - t) \Xi} b_{\rm GEM,in}(t), \label{gemsolna} \\
b_{\rm GEM,out}(T + t) = \chi(\zeta) \int_{-\infty}^\infty d\xi e^{-i xi t - i \zeta \ln t \Xi}  a(\xi,T), \label{gemsolnb}
\end{align}
where $\chi(\zeta) = \sqrt{\zeta} e^{-\pi \zeta/2}/\Gamma(1 - i\zeta)$, which has the property $|\chi(\zeta)|^2 = (1 - e^{-2\pi \zeta})/2\pi$. Strictly speaking \eref{gemsolna} and \eref{gemsolnb} describe the wavepackets of the state, rather than the operators, which we consider $a$ and $b$ to mean from this point on (see \cite{hush_analysis_2013} for details). We can see that what is stored in the memory is almost the Fourier transform of the input pulse, up to a time dependent phase distortion which is independent of the input pulse. If we pass the output of the GEM through a time dependent phase plate, i.e. $b'_{\rm GEM,out}(T + t) = S_{PP}(T+t) b_{\rm GEM,out}(T + t)$ where $S_{PP}(T+t) = 1$ for $t<0$ and $e^{i \zeta \ln (t (T-t) \Xi^2) - 2i \arg( \chi(\zeta))}$ for $t \ge 0$, then the final output of the memory can be related to the input as:
\begin{align}
b'_{\rm GEM,out}(T + t) = & (1 - e^{-2\pi \zeta}) b_{\rm GEM,in}(T - t). \label{eqn:gemfsoln}
\end{align}
Hence the output of the GEM is a time reversed, attenuated version of the input. The attenuation can be made arbitrarily small by increasing the optical depth $\zeta$. This makes the GEM the ideal candidate to reverse a quantum channel. 

\section{State transfer between two ensembles of rare earth ions} \label{sec:reitrans}

We propose an specific implementation of the generic state transfer using rare earth ion crystals as shown in \fref{fig:diagram}. An emitter crystal is coupled to an optical quantum channel through a mediating ring cavity. The output of the emitter is fed into a GEM and stored. The gradient of the GEM is flipped and the stored light is sent to the receiver crystal. The GEM acts as a perfect impedance matcher for the receiver ensemble by time reversing the light it received from the emitter ensemble. 

\subsection{The rare earth ion emitter and receiver ensembles}

We assume that two spectrally identical ensembles of ions in the emitter and receiver crystals have been prepared in the same hyperfine state, which we label $|g\rangle$, using laser pumping techniques. To begin, we neglect the inhomogeneous linewidth of the ensemble. This assumption is not necessary, which we show at the end of this section; however, we make it to ensure we can compare the dynamics of a single ion vs an ensemble on equal footing (the inhomogeneous linewidth is inherently a property of an ensemble). We assume an electric field can be applied to bring an optical transition between the ground state $|g \rangle$ and some excited state $|e \rangle$ into resonance with the cavity and driving laser. We also employ so-called cycling transitions \cite{bartholomew_engineering_2016} by applying appropriate magnetic fields to minimize spontaneous emission events into other unwanted hyperfine levels. For example, Pr$^{3+}$:Y$_2$SiO$_5$ could be used, where the $^3H_4$ level is used for $|g\rangle$ and $^1D_2$ is used for $|e\rangle$ \cite{bartholomew_engineering_2016}.

The emitter and receiver ions are described by the following Hamiltonian and coupling operator to the quantum channel:
\begin{align}
H_{l} = & -is_{l} g (J_{l} c^\dag_l - J_{l}^\dag c_l), \\
L_{l} = & \sqrt{\kappa} c_{l}.
\end{align}
Here the cavity has been brought into resonance with the optical transition between $|e \rangle \rightarrow |g \rangle$ and we  have moved into a rotating frame. 
The ensemble label $l$ represents the emitter $l={\rm em}$ or receiver $l={\rm re}$, $g$ is the coupling strength between an individual ion and the cavity, $c_l$ is the cavity annihilation operator, $\kappa$ is the coupling rate between the cavity and the optical channel, and $s_{l}$ is the sign of the Hamiltonian: through phase matching the cavity to the ensemble we set $s_{\rm em}=1$ and $s_{\rm re} = -1$. $J_{l}$ is the collective coupling operator between the ions and the cavity. We assume the cavity mode is a plane wave with wave vector $\mv{k}_0$, such that the coupling operator is $J_l = \sum_{j=1}^{N_l} e^{-i \mv{x}_{l,j} \cdot \mv{k}_0} \sigma_{l,j}.$ Here $\sigma_{l,j} = |g \rangle \langle e|_{l,j}$, and $N_{l}$ is the number of ions in the emitter and receiver. Except in the last figure, we set the number of ions to be the same and simply refer to the number of both as $N=N_{\rm em}=N_{\rm re}$; $\mv{x}_{l,j}$ is the position of the $j$th ion in the emitter or receiver. The position of the ions will be fixed, so we can make a transformation of each two level system such that $e^{-i \mv{x}_{l,j} \cdot \mv{k}_0} \sigma_{l,j} \rightarrow \sigma_{l,j}$. The collective coupling operator becomes:
\begin{equation}
J_l =\sum_{j=1}^N \sigma_{l,j}. 
\end{equation}
Even if the positions of the ions are different in the emitter and receiver, after making the transformation, the Hamiltonians will now be identical up to a sign change.  We assume each individual ion also undergoes spontaneous emission loss at a rate $\gamma$, i.e. $L^{\rm loss}_{l, j} = \sqrt{\gamma} \sigma_{l,j}$. 

The emitter and receiver ensembles satisfy some of our requirements for perfect state transfer. They have identical coupling operators and their Hamiltonians are of opposite sign $H_{\rm em} = -H_{\rm re}$. However, the Hamiltonian and loss operators do not necessarily have a unique dark state when $N>2$. This breaks the unique dark state condition required for perfect state transfer as shown in \sref{sec:gentrans}. We discuss how to circumvent this issue by restricting ourselves to only initial states where the ions are identical in the following section.

\subsection{The state initial conditions} \label{sec:stateic}

We consider transferring the state of an ensemble of identical ions. Specifically we restrict the initial condition of the emitter ions and the cavity to be of the form $|\psi_{\theta_0, \phi_0} \rangle$ $=|0 \rangle \otimes_{j=1}^N (\sin (\theta_0/2) |g\rangle$ $+\cos (\theta_0/2) e^{i\phi_0} |e\rangle)_{{\rm em},j}$ where $|0\rangle$ is the vacuum state for the cavity. We are effectively encoding one quantum bit onto an ensemble of ions. There are three reasons for this choice. First, states of this kind are easy to prepare physically using optical laser pulses \cite{longdell_experimental_2004}. Second, we can now make a fair comparison between the transfer fidelity of an individual ion or an ensemble of ions, as there is a one-to-one mapping between initial conditions. Lastly, if we start with the state $|\psi_{\theta_0, \phi_0} \rangle$ the system will converge to a unique dark state, as shown below, satisfying the final requirement for perfect state transfer.

The Hamiltonian $H_{l}$ and coupling operator $L_{l}$ are invariant under permutations of the ions, which means there is a symmetry in our system that is preserved during the evolution. Consider the symmetric states of the ions, e.g. for $N=3$ there are four states that form a basis for the symmetric subspace: $\{ |ggg\rangle, $ $(|egg\rangle + |geg\rangle +|gge\rangle)/\sqrt{3},$ $(|gee\rangle + |ege\rangle + |eeg\rangle)/\sqrt{3},$ $|eee\rangle \}$. We label the basis states for the symmetric subspace as $|n,N \rangle$, where $N$ is the total number of ions and $n$ is between $0$ and $N$ and refers to the number of excitations $e$ in the state. Consider the subspace spanned by the symmetric ion states and the basis states for the cavity $\mathcal{H}_{\rm{sym},N}$ $= \rm{span}[\{|m\rangle \otimes |n,N \rangle $ $| m \in [0,\infty)$ and $n \in [0,N] \}]$.  One can show $H |\psi_1 \rangle = |\phi_2\rangle$ and $L |\psi_3 \rangle = |\psi_4 \rangle$, where $|\psi_1\rangle,$ $|\psi_2\rangle,$ $|\psi_3\rangle$ and $|\psi_4\rangle$ are in $\mathcal{H}_{\rm{sym},N}$, such that if the ion-cavity system's initial condition is in $\mathcal{H}_{\rm{sym},N}$, it will stay in $\mathcal{H}_{\rm{sym},N}$ as the system evolves.

Our initial condition for the system $|\psi_{\theta_0, \phi_0} \rangle$ is a member of the symmetric subspace $\mathcal{H}_{\rm{sym},N}$. Furthermore, one can show there is a unique dark state in $\mathcal{H}_{\rm{sym},N}$, specifically $|\psi_{\rm ss}\rangle = |0 \rangle \otimes |0, N \rangle$, as $L |\psi_{\rm ss}\rangle = 0$ and $H |\psi_{\rm ss}\rangle = 0$. Hence, for our restricted set of initial conditions, we can guarantee that our system will converge to the unique dark state. This is very important, as it means our system now satisfies all the conditions required for perfect state transfer (as shown in \sref{sec:gentrans}).

\subsection{The total system}

We have shown that the emitter and receiver, when restricted to an initial condition of $|\psi_{\theta_0, \phi_0} \rangle$, satisfy the requirements for perfect state transfer. The last step is to connect the emitter optical output and receiver optical input via a GEM, which will perform the required time reversal of the optical channel. The total Hamiltonian $H_{\rm total}$ and coupling operator $L_{\rm total}$ for the entire emitter-GEM-receiver system can be derived using input-output theory \cite{gardiner_quantum_2004,gough_series_2009}:
\begin{align}
H_{\rm total} = & H_{\rm em} + (L^\dag_{\rm GEM} L_{\rm em} + L_{\rm GEM} L_{\rm em}^\dag)/(2i)  + H_{\rm GEM} \nn \\ 
& + (L^\dag_{\rm re} L_{\rm GEM}' + L_{\rm re} {L_{\rm GEM}'}^\dag)/(2i) + H_{\rm rm}, \\
L_{\rm total} = & L_{\rm GEM}' + L_{\rm re},
\end{align}
where $L_{\rm GEM}' = S_{PP}(t)(L_{\rm em} + L_{\rm GEM})$. 

The timing of the transfer proceeds as follows: at $t=0$ the emitter crystal is prepared in some state $|\psi_{\rm em}(0) \rangle = |\psi_{\theta_0, \phi_0} \rangle$. The output of the emitter is written on to the GEM over a time period $T$, which is sufficient time for the emitter to enter the dark state $|\psi_{\rm em}(T) \rangle \approx |\psi_{\rm ss}\rangle$. The gradient of the GEM is flipped at time $T$ and the output is passed through the phase plate to the receiver ensemble. The receiver has been initialized in the dark state: $|\psi_{\rm re}(T) \rangle = |\psi_{ss} \rangle$. The time flipped output of the emitter is then absorbed by the receiver. In the ideal case, we expect the final state of the receiver to be the initial state of the emitter $|\psi_{\rm re}(2T) \rangle \approx |\psi_{\theta_0,\phi_0}\rangle$.

In what follows we perform a set of numerical simulations to verify the performance of the GEM in ideal conditions and investigate how loss affects the transfer performance.

\subsection{Coupling to the optical channel}

\begin{figure}[t!]
\centering
\includegraphics[width=1.0\linewidth]{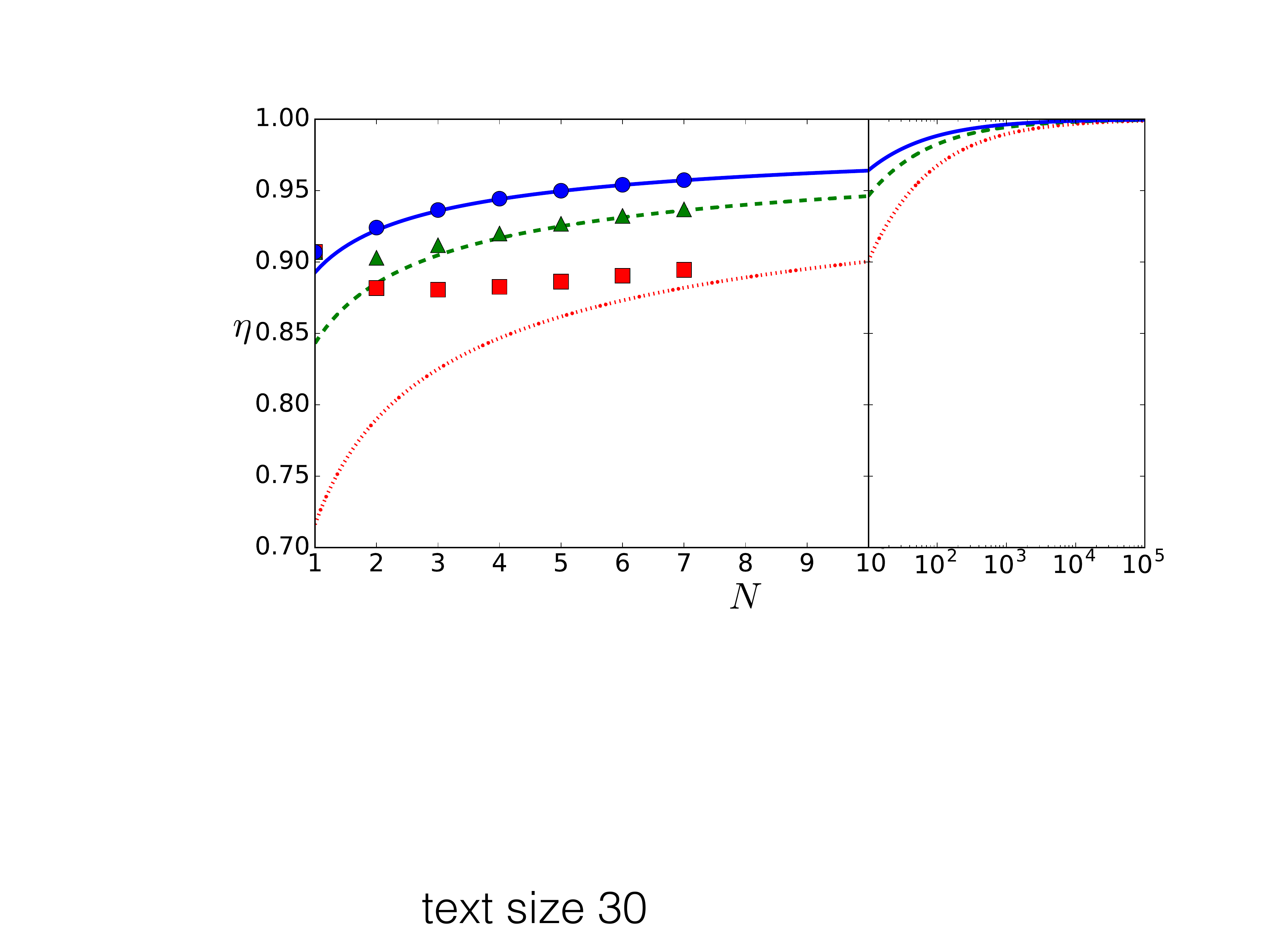}
\caption{The efficiency $\eta$ of coupling to the optical channel vs particle number $N$. Direct simulations are plotted as points, with initial conditions: $\theta_0 = \pi/4$ (red squares), $\pi/2$ (green triangles) and $3\pi/4$ (blue squares). Mean field simulations are are plotted as lines, with initial conditions:  $\theta_0 = \pi/4$ (dotted red line), $\pi/2$ (dashed green line) and $3\pi/4$ (solid blue line). Here $\gamma/g = 0.1$, $\kappa = 2g\sqrt{N}$ and $T = 20/\kappa$.}
\label{fig:outeff}
\end{figure}

\begin{figure*}[t!]
\centering
\includegraphics[width=0.7\linewidth]{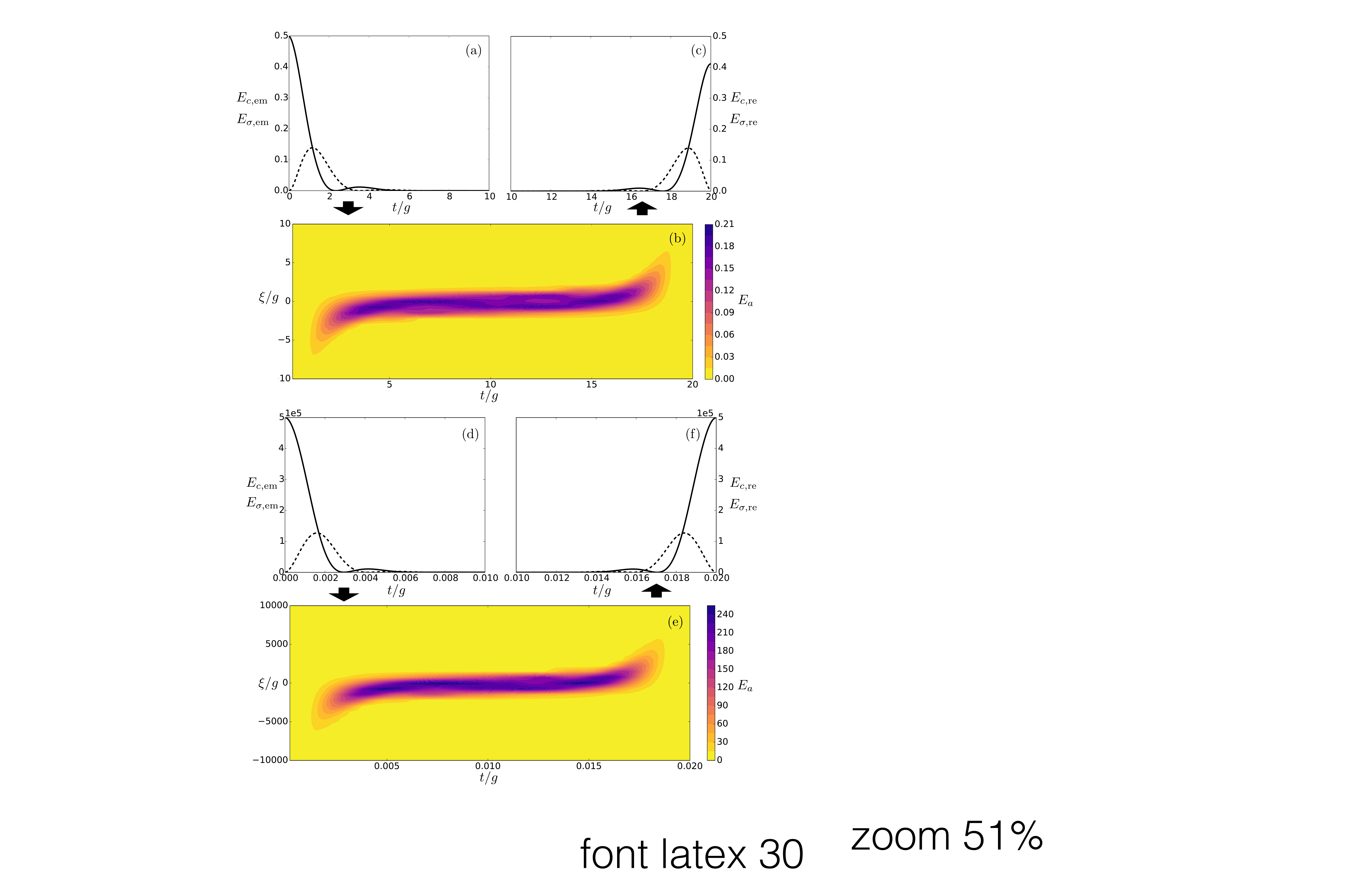}
\caption{Dynamical flow of excitation between emitter $\rightarrow$ GEM $\rightarrow$ receiver vs time $t$ with $\gamma/g = 0.1$, $\zeta = 2$, $\theta_0 = \pi/2$, $\phi_0=0$, $\kappa/g = 2\sqrt{N}$ and $T = 20/\kappa$. Direct simulations are presented in (a)-(c) with $N=1$ and mean field results are presented in (d)-(f) with $N=10^6$. In subfigures (a), (c), (d) and (f) the solid line is the excitation of the ions $E_{\sigma,l}(t)$ while the dotted line is the excitation of the cavity $E_{c,l}(t)$. Subfigures (b) and (e) are plots of the excitation in the memory $E_{a}(\xi,t)$, plotted also against frequency $\xi$. The arrows indicate the movement of excitation. The final fidelity of transfer for the $N=1$ simulation was $\mycal{F} = 95.35\%$ while the $N=10^6$ simulation was $\mycal{F} = 99.98\%$.}
\label{fig:transevo}
\end{figure*}

\begin{figure*}[t!]
\centering
\includegraphics[width=0.7\linewidth]{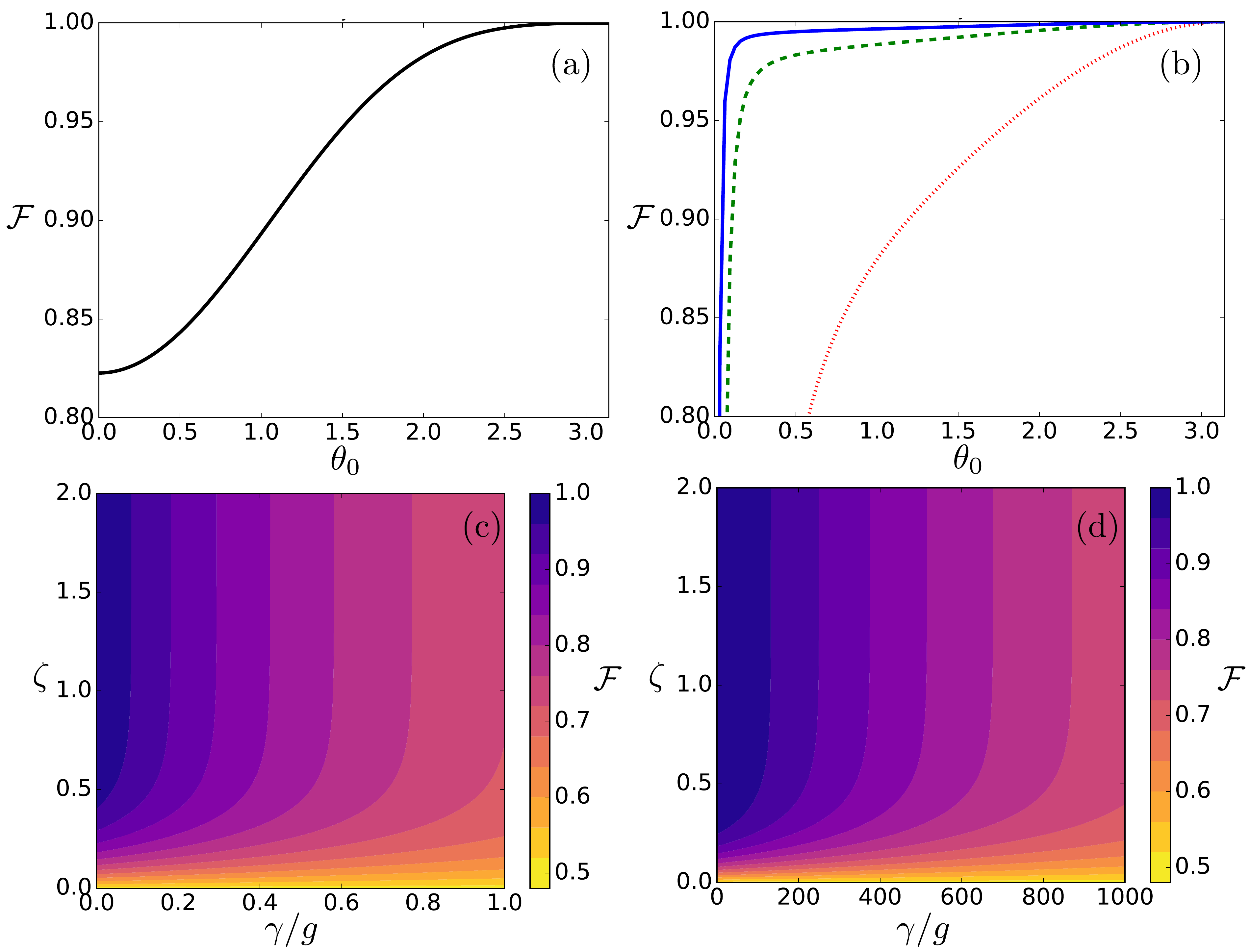}
\caption{Fidelity $\mycal{F}$ of the transfer compared to different system parameters. The direct method is used for (a) and (c) with $N=1$ while the mean field method is used for (b) and (d) with $N \gg 1$. Fidelity is plotted against the initial state $\theta_0$ in (a) with $N=1$ (solid black line), and in (b) with $N=10$ (dotted red line), $N=10^3$ (dashed green line) and $N=10^4$ (solid blue line); in both plots $\gamma/g=0.1$ and $\zeta = 2$. Fidelity is plotted against the optical depth $\zeta$ and loss $\gamma$ in (c) and (d) with $\theta_0 = \pi/2$, $N=1$ in (c) and $N=10^6$ in (d). All simulations have $\phi_0=0$, $\kappa = 2g\sqrt{N}$, and $T = 50/\kappa$.}
\label{fig:transfid}
\end{figure*}

Before we consider the entire transfer process, we consider the possible advantages of using an ensemble rather than individual rare earth ions when coupling to a optical channel. 

Our first challenge is to engineer the emitter such that the majority of excitation is transferred into the optical channel instead of other loss mechanisms.
We have assumed that each individual ion undergoes spontaneous emission, with rate $\gamma$, which is not captured by our optical channel. 
We thus want to couple the light from the cavity into the optical channel as fast as possible. By linearizing the ensemble it can be shown that the critical damping rate for the cavity is $\kappa = 2g\sqrt{N}$, which we use for the remainder of the paper. 

Next we look at the parameters of the ions. There are $N$ ions which are collectively coupled to the cavity with a strength $g$. This results in a superradiant enhancement of the effective coupling strength to the optical channel. We do not use this approximation in our simulations, but for the purpose of analysis, we can adiabatically eliminate the cavity, which gives an effective coupling operator $L_{\rm em}' = 2g J_{\rm em}/\sqrt{\kappa}$ between the ensemble and the optical channel. In this limit, we can see that the coupling between the ions and the channel is collective. This results in $N^{3/2}$ scaling of the spontaneous emission rate into the optical channel, namely, $P_{col} = \langle \psi_{\theta_0, \phi_0} | L_{\rm em}'^\dag L_{\rm em}' |\psi_{\theta_0, \phi_0} \rangle = 2N^{3/2} g \cos^2 \theta_0 $, where we have replaced $\kappa$ with its critical damping value. 

In contrast, the probability of an independent spontaneous emission event into one of the loss channels scales as $N$, namely $P_{loss} = \sum_{j=1}^N \langle \psi_{\theta_0, \phi_0} | (L^{\rm loss}_{{\rm em}, j})^\dag L^{\rm loss}_{{\rm em}, j} |\psi_{\theta_0, \phi_0} \rangle = N \gamma \cos^2 \theta_0$.

We are interested in the relative rate of spontaneous emission into the optical channel compared to other modes, specifically $R = P_{col}/P_{loss} = 2 \sqrt{N} g/\gamma$.  

Our aim is to make $R$ as large as possible, to maximize the light spontaneously emitted into the optical channel compared to other loss channels. Increasing $g/\gamma$ is possible by using cavities with small mode volumes \cite{mcauslan_strong-coupling_2009}, but achieving this in practice with rare earth ions has been challenging \cite{mcauslan_coherent_2011}. In contrast, achieving extremely large $N$ in rare earth ions is straightforward. A large $R$ is the key advantage ensembles have over individual rare earth ions. 

We demonstrate the advantage of large $N$ for the coupling efficiency $\eta = I_{out}/N \cos^2 (\theta_0/2)$ numerically in \fref{fig:outeff}.  The coupling efficiency is defined as the total light emitted into the optical channel $I_{out} = \int_{0}^T \langle b_{out}^\dag(t) b_{out}(t) \rangle$ compared to the total excitation in the initial ensemble of atoms $N \cos^2 (\theta_0/2)$. We perform the simulation with a direct method for $N=1$ to $7$ and a mean field method for larger particle numbers, see \aref{app:effcoup} for details. Furthermore, we consider three initial states of $\theta_0$: $\pi/4,\pi/2$ and $3\pi/4$. In all cases the quantum efficiency eventually improves with particle number, although there is a small dip in the direct simulations after $N=1$.

In terms of the convergence between the mean field method and direct method, we see that the solutions asymptotically approach one another in the limit of large $N$. However, the rate of convergence is slower for states close to $\theta_0 = 0$. The convergence is so slow in the $\theta_0 = \pi/4$ case that we can not confirm, quantitatively, that the methods converge with the maximum number of particles we could simulate with the direct method: $N=7$. The reason for this slow convergence is that the mean-field description of the coupling has a nonphysical, unstable fixed point for an initial condition of $\theta_0 = 0$ (see \aref{app:effcoup}). This means the mean-field simulations tend to significantly underestimate the true efficiency for initial states close to the excited state. Fortunately, as we can see in \fref{fig:outeff}, the mean-field is always strictly below the direct solution. Hence it can be thought of as a lower bound for the efficiency that gets tighter as the particle number $N$ increases.

The nonphysical unstable fixed point present in the mean-field equations of motion means we cannot look at the $\theta_0 = 0$ case directly. In what follows we investigate scaling as the state gets close to $\theta_0 = 0$. Strictly speaking, this limitation means the collective effects we see in the state transfer will not include all stages of a superradiant process. Our analysis only considers the second stage in a superradiant process where a small perturbation from a fully inverted population is rapidly damped to the ground state. We do not model the initial spontaneous emission event that perturbs the ensemble from the fully inverted stage \cite{macgillivray_theory_1976,polder_superfluorescence:_1979}. It would be interesting to probe this regime in an experiment.

\subsection{Complete quantum state transfer}

We perform a numerical simulation of the quantum state transfer protocol in a system with small loss, $\gamma/g = 0.1$, and a high quality GEM with optical depth $\zeta = 2$. Direct simulations of a GEM become rapidly computationally expensive as the number of photons increases. Fortunately, as we have demonstrated in \fref{fig:outeff}, mean-field simulations make reliable quantitative predictions for ensembles with $N \ge 7$ and initial states $\theta_0 \ge \pi/2$, and give a good lower bound on fidelity for initial states $\theta_0 < \pi/2$ that tightens as $N$ increases. Hence, we use two simulation methods: a direct method for the $N=1$ case, and a mean-field method for $N \gg 1$ (see \aref{app:fidtrans} for details).

In \fref{fig:transevo} we plot the dynamical flow of excitation from the emitter to the GEM, then finally to the receiver. We define the excitation for: the ions as $E_{\sigma,l}(t) =  \sum_{j=1}^N  \langle \sigma^\dag_{l,j} \sigma_{l,j} \rangle$, cavities as $E_{c,l}(t) = \langle c^\dag_l c_l\rangle$, and the memory as $E_{a}(\xi,t) = \langle a(\xi)^\dag a(\xi) \rangle$. In the $N=1$ case, we can see that the evolution of the emitter and receiver is close to symmetric about $T$, and the receiver almost reaches the initial state of the emitter. The finite amount of loss $\gamma/a$ results in some loss of excitation and an imperfect transfer. However, this loss can be overcome through superradiant coupling. This is shown in the $N=10^6$ case. Here we see that the evolution of the emitter and receiver is now perfectly symmetric and the receiver finishes almost exactly in the initial state of the emitter. Furthermore, comparing the ensemble to individual ion transfer, we can see that the shape of the cavity output does depend on $N$, and the time it takes to perform the transfer is much shorter in the $N=10^6$ case compared to the $N=1$ case.

\subsection{Effect of loss}

To get a more detailed understanding of how imperfection affects the system, we now look at the transfer fidelity as a function of initial state, number, loss and optical depth in \fref{fig:transfid}. We define the fidelity $\mycal{F}$ of the transfer with regard to the average state of the ion. Specifically, we define the average ion state to be $\bar{\rho}_{l} = \sum_{j=1}^N \Tr_{l,j}[\rho]/N$ where $\Tr_{l,j}$ is defined as tracing over all systems \emph{except} the $j$th ion of the emitter $l= {\rm em}$ or receiver $l={\rm re}$ and the fidelity is $\mycal{F} = ||\sqrt{\bar{\rho}_{\rm em}(0)} \sqrt{\bar{\rho}_{\rm re}(2T)}||$ where $|| \cdot||$ is the trace norm. 

In \fref{fig:transfid}(a) we consider how the initial state affects the transfer process. For a fixed $\gamma$ and $\zeta$, we plot the fidelity of state transfer as a function of the initial state. The state transfer fidelity only depends on the excited state population, and is independent of $\phi_0$. The transfer fidelity clearly depends on $\theta_0$ and is worst when the state is initialized in $|\psi_{\rm em}(0) \rangle = |e\rangle$. This is because our protocol transfers the amplitude of the excited state, but the receiver is already initialized in the ground state. Consequently,  we will always achieve a perfect transfer fidelity when $|\psi_{\rm em}(0) \rangle = |g\rangle$, independent of any imperfections in the system.

Furthermore, we see that the lower-bound provided by the mean field prediction is overly conservative when the states are close to $\theta_0=0$. As we increase the number of particles $N$, the lower bound on fidelity improves. However, the fixed point at $\theta_0=0$ makes the mean-field simulation technique uninformative for the $\theta_0=0$ case, and we have to analyze the situation physically.

When $\theta_0=0$, there exists no relative phase relationship between the ions in the ensemble as they are all in the excited state. In this case, the probability of the \emph{first} emission into the cavity mode compared to other modes will be independent of the particle number. Instead it will only be a function of the geometry, or more specifically, the mode volume of the ensemble. In this case the ensemble will not necessarily have an advantage over an individual ion. But this is only true when $\theta_0=0$; when $\theta_0 = \epsilon$ is small, a phase relationship will develop between the ions, meaning emission into the cavity vs other modes will start to scale with $N$. More specifically, we expect an ensemble to start having a distinct advantage over an individual ion (for the same mode volume) when $\epsilon > 1/\sqrt{N}$. Thus the volume of initial states where an ensemble has a clear advantage over an individual ion also increases with particle number. Lastly, working with an ensemble provides more flexibility with geometry, which may provide an advantage over an individual ion even in the case of $\theta_0 = 0$, but we leave this as an open question for future work.

In \fref{fig:transfid}(b)-(d) we consider the deleterious effects of loss and lower optical depths. We perform a scan over loss $\gamma$ and optical depth $\zeta$ versus fidelity, with a fixed $g$ and initial state. We can see that the fidelity monotonically decreases as the loss increases, or when the optical depth gets smaller. In most applications there will be some finite fidelity for transfer required before error correction can be employed to compensate. Fortunately, given some initial state and target fidelity, we can overcome a finite loss rate $\gamma$ by simply using a larger $N$, as shown in \fref{fig:transfid}. Similarly, the performance of a GEM can be improved by using a larger optical depth $\zeta$, which can be achieved by increasing the rare earth ion density. In both cases, stoichiometric rare earth ion crystals with narrow inhomogeneous linewidths could be used to achieve very large $N$ and $\zeta$ as required for highly efficient state transfer.

\subsection{Effect of number imbalance}

\begin{figure}[t!]
\centering
\includegraphics[width=0.7\linewidth]{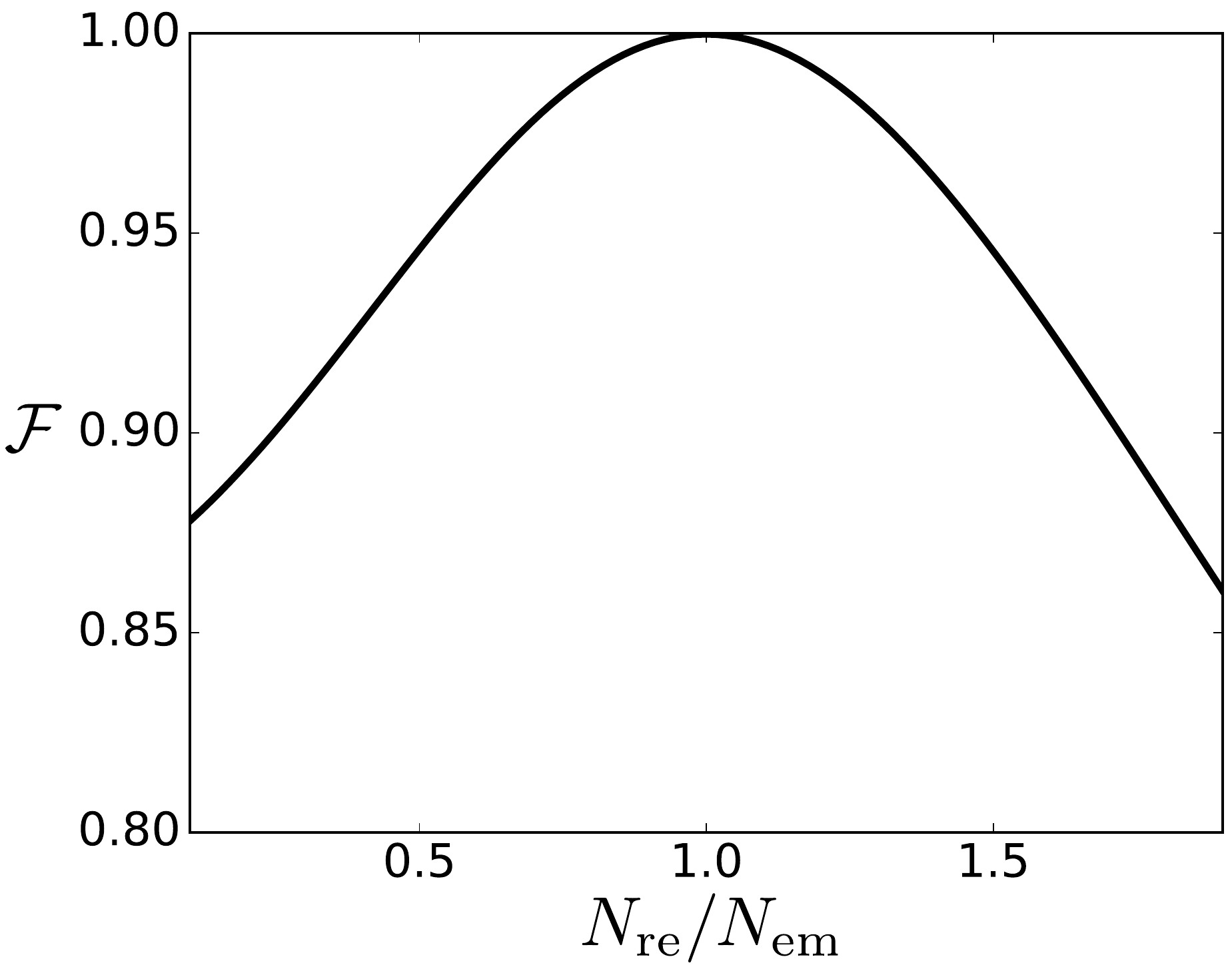}
\caption{Fidelity $\mycal{F}$ of the transfer compared to the ratio between the number of ions in the receiver compared to emitter $N_{\rm re}/N_{\rm em}$ with $N_{\rm em} = 10^6$, $\theta_0 = \pi/2$, $\phi_0 = 0$, $\zeta = 2$, $\gamma = 0$, $\kappa = 2g\sqrt{N_{\rm em}}$ and $T = 50/\kappa$.}
\label{fig:nimb}
\end{figure}

When working with ensembles, another issue that can occur is an imbalance in the number of ions between the emitter and receiver. We can only numerically investigate this issue with the mean-field method. In \fref{fig:nimb} we plot the fidelity of transfer as a function of the ratio $N_{\rm re}/N_{\rm em}$. We can see the transfer fidelity is perfect when $N_{\rm em} = N_{\rm re}$, but reduces as the ratio of ions in the emitter and receiver becomes unbalanced. Fortunately, this problem is solved by simply using a larger ensemble. Assuming that the numbers of ions in the ensembles are randomly chosen from a Poisson distribution with the same mean $N$, which is reasonable given they are typically prepared optically with lasers that also obey Poisson statistics, the standard deviation in the distribution of the ions will scale as $\sqrt{N}$. Given this distribution, the mean fraction will be $N_{\rm re}/N_{\rm em} = 1$, and the standard deviation in the fraction will be $1/\sqrt{N}$. Consequently, using a larger ensemble results in a fraction close to one, and a higher fidelity. 

\subsection{Effect of inhomogeneous broadening}

\begin{figure*}[t]
\centering
\includegraphics[width=0.7\linewidth]{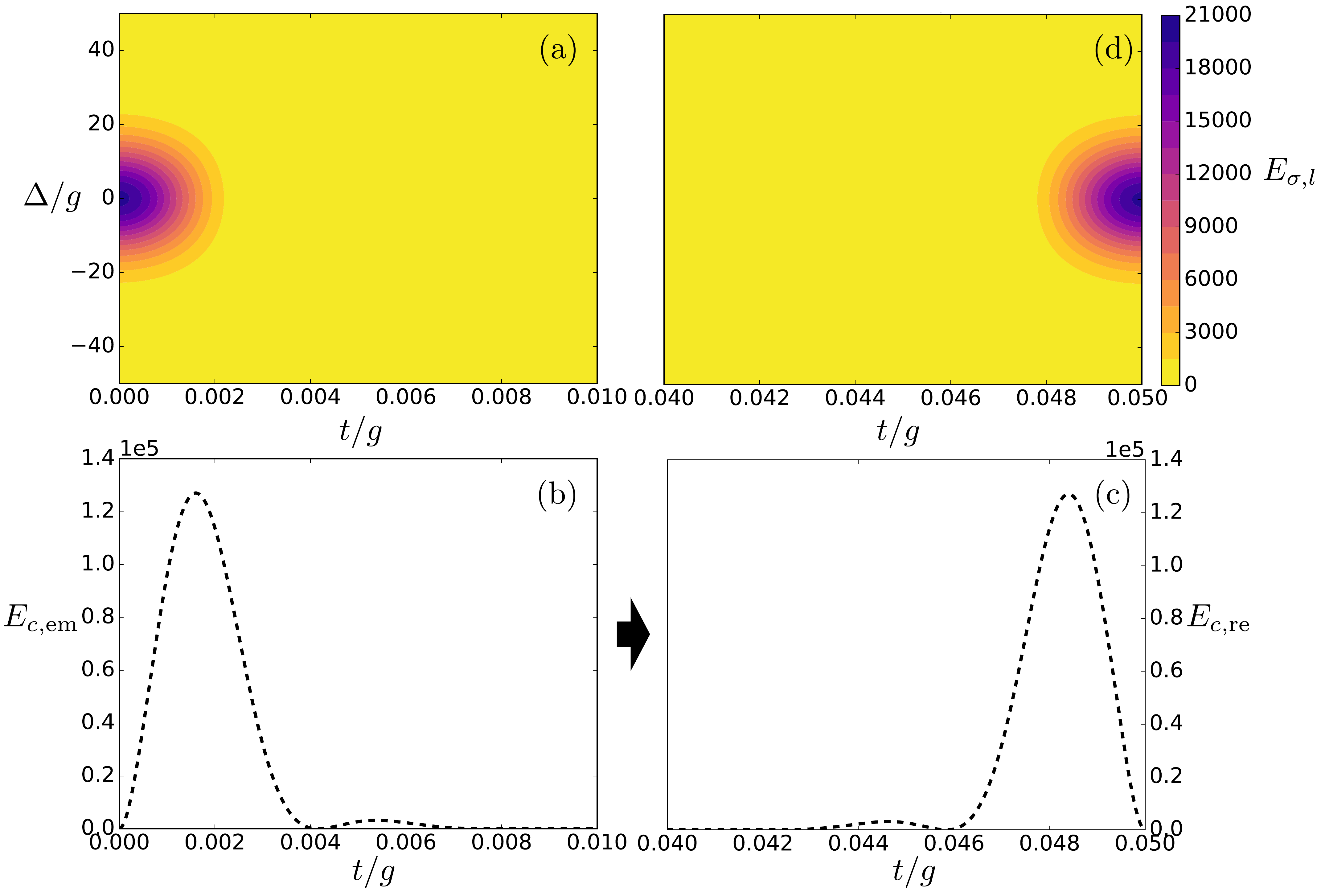}
\caption{Dynamics of excitation during state transfer with inhomogeneous broadening. In subfigures (a) and (d) the excitation spectral density $E_{\sigma,l}(\Delta,t)$ of the ions is plotted against the detuning $\Delta$ and time $t$ in the emitter and receiver ensemble respectively. In subfigures (b) and (c) the excitation in the cavity is plotted against time withe a dashed line in the emitter and receiver respectively. The excitation in the GEM is not shown. The parameters are the same as the mean field simulations presented in \fref{fig:transevo} with $N=10^6$, $\gamma/g = 0.1$, $\zeta = 2$, $\theta_0 = \pi/2$, $\phi_0 = 0$, $\kappa/g = 2\sqrt{N}=10^3$ and $T= 50/\kappa$, except we now include inhomogeneous broadening with width $\sigma_{\Delta}/g = 10$. The final fidelity of transfer is $\mathcal{F} = 99.95\% $}
\label{fig:finhomo}
\end{figure*}

In practice, the Hamiltonians for the emitter and receiver ensemble will include an additional term to account for the inhomogeneous broadening of the ensemble: 
\begin{align}
H_{l} = & \sum_{j=1}^N \Delta_{l,j} \sigma_{l,j}^{\rm z} -is_{l} g (J_{l} c^\dag_l - J_{l}^\dag c_l). \label{eqn:hamilinhomo}
\end{align}
Each $\Delta_{l,j}$ is a random variable sampled from the inhomogeneous broadening density function $\varrho_{l}(\Delta)$, with $\int_{-\infty}^{\infty} d\Delta \varrho_{l}(\Delta) = N$. We assume that the inhomogeneous broadening distribution is the same for the emitter and receiver $\varrho(\Delta) =\varrho_{\rm em}(\Delta) = \varrho_{\rm re}(\Delta)$ and it is an even function such that $\varrho(\Delta) = \varrho(-\Delta)$. In the limit of very large $N$, $H_{\rm em} \approx -H_{\rm re}$. There still exists a unique dark state where all the ions are in the ground state. Consequently, even with the addition of inhomogeneous broadening, efficient state transfer is possible. 

We numerically investigate state transfer with an inhomogeneous linewidth in \fref{fig:finhomo}, where we plot the flow of excitation during the transfer process, with the same parameters as \fref{fig:transevo}, but we now include significant broadening. The equations of motion we used for the simulations, which were derived by applying a mean field approximation and going to the continuum limit, are presented in \aref{app:inhomotrans}. We set $\varrho_{l}(\Delta) = N e^{-\Delta^2/2\sigma_{\Delta}^2}/ \sigma_{\Delta}\sqrt{2\pi}$ with $\sigma_{\Delta}/g = 10 $; this linewidth is much larger than both the coupling strength $g$ and the homogeneous linewidth $\gamma/g=0.1$, but it is still much smaller than the cavity damping $\kappa/g=2\sqrt{N}=2000$. The transfer fidelity is $99.95\%$, which is slightly less than the case without broadening, because the superradiant amplification of the emission is slightly reduced by dephasing caused by the inhomogeneous linewidth. Nevertheless, this transfer fidelity is still high and can be further improved by increasing $N$. 

\begin{figure}[t!]
\centering
\includegraphics[width=0.75\linewidth]{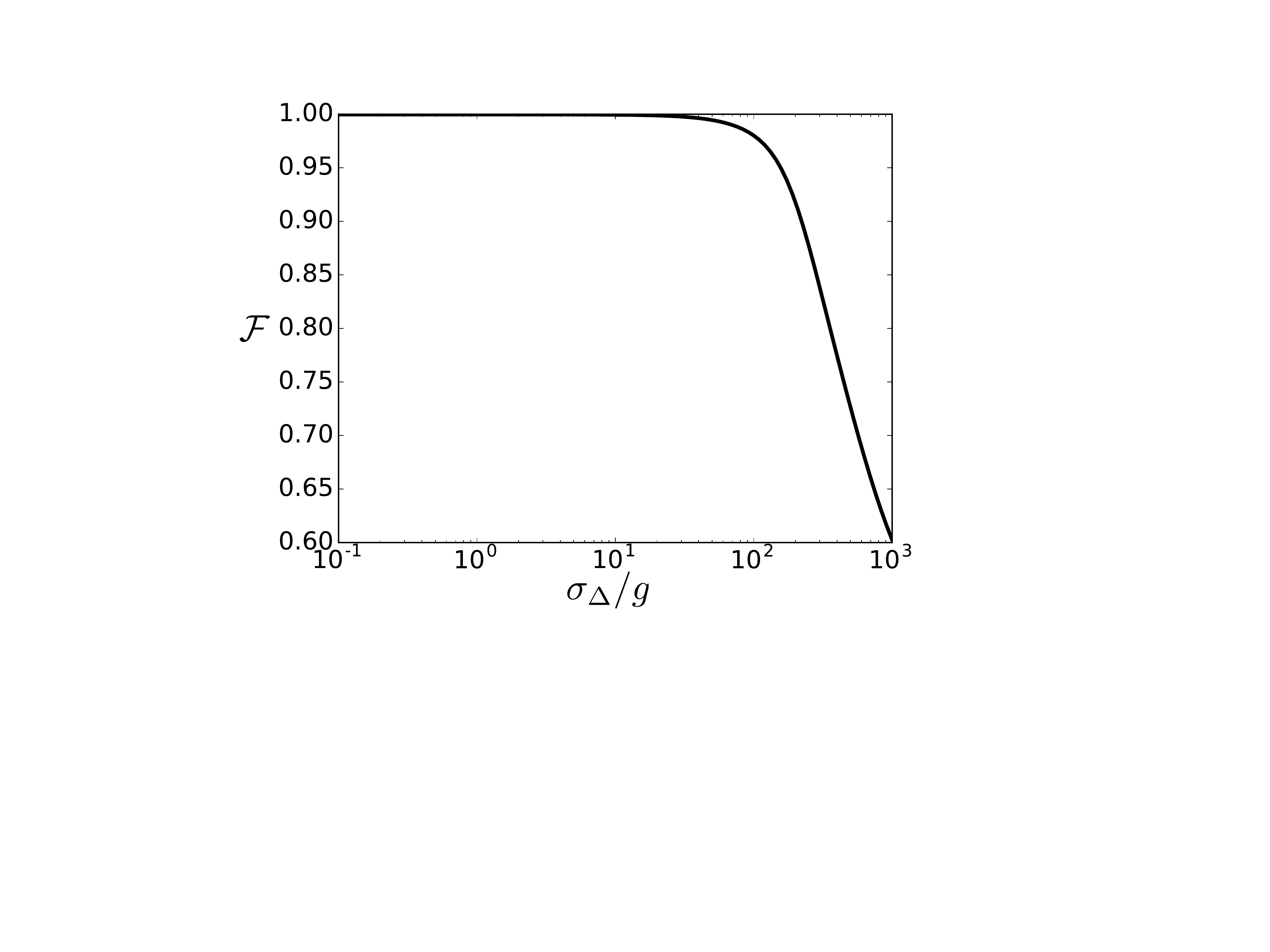}
\caption{Fidelity $\mycal{F}$ of the transfer compared to the inhomogeneous broadening $\sigma_{\Delta}$ with $N = 10^6$, $\theta_0 = \pi/2$, $\phi_0 = 0$, $\zeta = 2$, $\gamma/g = 0.1$, $\kappa/g = 2\sqrt{N}= 10^3$ and $T = 50/\kappa$. Note that the fidelity does not significantly drop until $\sigma_{\Delta}$ becomes comparable to $\kappa$.}
\label{fig:finhomoscan}
\end{figure}

In \fref{fig:finhomoscan} we present a plot of the transfer fidelity $\mathcal{F}$ as a function of $\sigma_{\Delta}$. Here we can see the fidelity is only significantly affected by the linewidth $\sigma_{\Delta}$ when it becomes large enough that it is comparable to $\kappa$. When $\sigma_{\Delta}$ is similar to $\kappa$ the ions will begin to dephase before the excitation has had time to escape through the cavity. This leaves some excitation effectively trapped in the ions, which is not transferred. This issue can be solved in two ways. First, a larger spectral density $N$ can be used, which will make $\kappa$ larger and ensure the excitation leaves the system before significant dephasing has occurred. Second, a sequence of pi-pulses could be used to stop the ions from dephasing and ensure that all the excitation is released from the system. In either case a large inhomogeneous linewidth $\sigma_{\Delta}$, which is bigger than $g$ and $\gamma$, is not a fundamental issue for efficient state transfer. 

\section{Transferring Entangled States} \label{sec:entangledstates}

Our discussion of state transfer has so far only considered sending an individual qubit of information that is separable from its environment. However, in quantum information processing it is important to transfer a qubit which may be entangled with some auxiliary system. It is straightforward to consider the case of a single ion entangled to an ancilla, which is discussed below, but generalizing to the ensemble case is complicated by the many types of possible entanglement between ensembles. It is beyond the scope of this manuscript to perform simulations of entangled ensemble state transfer; we discuss where we expect our results apply and where further engineering is required.

First we emphasize that our demonstration that generic state transfer can be performed through time reversal of an optical channel, in \sref{sec:gentrans}, was linear with regard to the quantum state. Hence, according to the superposition principle, as long as each of the separable states, that add up to form the entangled state, satisfy the conditions in \sref{sec:gentrans} the state transfer will work perfectly. For example, given some entangled state $(|\psi_1 \rangle \otimes |\phi_1\rangle +$ $|\psi_2\rangle \otimes |\phi_2\rangle)/\sqrt{2}$, if you can show the transfer will work perfectly for $|\phi_1\rangle$ and $|\phi_2\rangle$ it is guaranteed to work for the entangled state. 

Our method of state transfer will work with entangled states for the individual ion case $N=1$. For example, consider an entangled state $(|ee\rangle + |gg\rangle)/\sqrt{2}$ between two ions in two separate crystals and cavities. If we wish to transfer the state of the first ion, the transfer protocol works for both $|g\rangle$ and $|e\rangle$, hence, by the superposition principle, we can be confident it will work for $(|ee\rangle + |gg\rangle)/\sqrt{2}$. The fidelity is a nonlinear function, so the transfer fidelity for an entangled state will not be a simple function of the fidelities of the separable states, nevertheless, we can be confident the transfer fidelity will approach one as the system approaches the ideal case. 

Our method of state transfer will also work with ensembles, as long as the entanglement is between states in the symmetric subspace. As we discussed in \sref{sec:stateic}, our method works for ensembles because the initial condition is a symmetric state, the dynamics preserve this symmetry, and there is a unique dark state in the symmetric subspace. For example, if we had two ensembles of ions in separate cavities that were in the entangled state $(|\psi_{0, \pi/2} \rangle \otimes |\psi_{\pi, \pi/2} \rangle$ $+|\psi_{\pi, \pi/2} \rangle \otimes |\psi_{0, \pi/2} \rangle)/\sqrt{2}$, both $|\psi_{\pi, \pi/2} \rangle$ and $|\psi_{0, \pi/2} \rangle$ are in the symmetric subspace, hence perfect state transfer will be possible in an ideal system. However, some entangled states may be very sensitive to loss, for example consider the state  $(|\psi_{0, \pi} \rangle \otimes |\psi_{0, 0} \rangle$ $+|\psi_{0, 0} \rangle \otimes |\psi_{0, \pi} \rangle)/\sqrt{2}$, which is equivalent to a NOON state $(|N,0\rangle$ $+|0,N \rangle)/\sqrt{2}$ that will collapse to a separable state after the loss of a single photon. This suggests that the state transfer fidelity may drop rapidly as the system becomes less than ideal. An analysis of which entangled states are more or less sensitive to loss during the state transfer could be considered in future work. 

There are other entangled state where our approach needs further engineering and consideration. For example, consider an ensemble of entangled pairs of ions (labeled $a$ and $b$) in a single crystal: $\otimes_{j=1}^N (|g\rangle_a |g\rangle_b$ $+|e\rangle_a |e\rangle_b)_{j}$, where there are $2N$ ions in total. If we attempt to transfer the state of just one ion in each pair, we find, in many cases, that the states in the superposition are not in the symmetric subspace. For example, consider two entangled pairs $((|g\rangle_a |g\rangle_b+|e\rangle_a |e\rangle_b)_{1}$ $ \otimes (|g\rangle_a |g\rangle_b + |e\rangle_a |e\rangle_b)_{2})/2$. We can rearrange this state as $(|gg\rangle_a \otimes |gg\rangle_b$ $+(|ge\rangle + |eg\rangle)_a \otimes (|ge\rangle + |eg\rangle)_b /2$ $+(|ge\rangle - |eg \rangle)_a \otimes (|ge\rangle - |eg\rangle)_b /2$ $+|ee\rangle_a \otimes |ee\rangle_b)/2$. Note that three terms in this superposition are in the symmetric subspace for two ions spanned by $\{|gg\rangle,$ $(|ge\rangle + |eg\rangle)/\sqrt{2},$ $|ee\rangle\}$ and will be transferred. However, $(|ge\rangle - |eg\rangle)/\sqrt{2}$ is an anti-symmetric state, which is a dark state with respect to $H_{l}$ and $L_{l}$, and will not be transferred. Thus, even in the ideal case, there are limitations to our protocol for ensembles of entangled pairs. One way of circumventing this issue is to include the inhomogeneous broadening in the Hamiltonian. In this case the anti-symmetric state $(|ge\rangle - |eg\rangle)/\sqrt{2}$ will no longer be dark and should be transferred. However, we have shown that inhomogeneous broadening can suppress superradiant effects, and possibly reduce the advantages of an ensemble. Understanding this trade-off, or engineering other solutions, could be examined in future work.

From the brief discussion above we can see that state transfer of entangled ensembles will depend on the nature of the entanglement. There exist some entangled ensembles that are not entirely in the symmetric state subspace, and hence break the unique dark state condition required for the perfect state transfer. The focus of this manuscript has been investigating state transfer when the emitter and receiver have a unique dark state. Examining how to approach state transfer when this is not the case is a matter for future work.

\section{Discussion}  \label{sec:discus}
We have described a method for transferring a quantum state from one system to another by way of a GEM and described its implementation using rare earth ions. The implementation we presented includes an  optical coupling link between crystals to allow  for transport over long distances. However, this is not vital, and initial experimental demonstrations could even be made with a single crystal performing the role of emitter, memory, and receiver by using controllable electrodes to create three distinguishable regions along the direction of light. Likewise, the cavity is not the only way to enhance the superradiant coupling along the light direction. Another option is to change the geometry of the ensemble to make a long, skinny cylinder \cite{beavan_demonstration_2012}. This ensures superradiant amplification mostly occurs along the cylinder axis. With these modifications, implementing this protocol in a rare earth crystal is straightforward, particularly as GEMs have previously been demonstrated in rare earth crystals \cite{alexander_photon_2006}. The only additional component required in our implementation is the time-dependent phase shift at the output, which can be achieved using a time-dependent applied electric field to change the detuning of the rare-earth ions at the end of the GEM. 

In the broader context of quantum control, this work is a novel example of using a coherent non-causal filter to achieve a control goal. As we have performed a time reversal on the optical channel, the input field to the receiver is both non-Markov and non-causal \cite{gardiner_quantum_2004}. There has been extensive work on solving causal filtering problems with coherent quantum components \cite{james_control_2008,nurdin_coherent_2009,miao_quantum_2012,yamamoto_zero-dynamics_2014,miao_coherently_2015}, but very little on coherent non-causal filters. This is primarily because it has been unclear how to coherently implement a non-causal filter. Here we see GEMs are an excellent candidate to perform coherent non-causal filtering of a signal. More complex non-causal filters could be produced by reading out different parts of the memory at different times depending on the control goal.

In summary, we have shown generic state transfer is possible by time-reversal of a quantum optical channel. We have given an implementation of this protocol using rare-earth ion crystals and a GEM. Furthermore, we have demonstrated that state transfer of rare-earth ion ensembles is possible, and that the transfer fidelity of ensembles can benefit from collective phenomena, namely, superradiance. Lastly, we discussed where we expect our approach will work with entangled states, and where further engineering and understanding is required. 

\begin{acknowledgments}
MRH acknowledges funding from an Australian Research Council (ARC) Discovery Project (project number DP140101779). The authors thank JJ Hope and ARR Carvalho for helpful discussions. 
\end{acknowledgments}

\appendix

\section{Simulations for coupling to the channel} \label{app:effcoup}

Here we describe the simulation methods and approximations used to provide insight into the coupling between the emitter and optical channel. We only need to model the emitter in this case. The master equation for the emitter is:
\begin{align}
\frac{d\rho_{\rm em}(t)}{dt} = & -i [H_{\rm em},\rho_{\rm em}] + \mathcal{D}[L_{\rm em}](\rho) \nn \\
& + \sum_{j=1}^N \mathcal{D}[L^{\rm loss}_{\rm em,j}](\rho).
\end{align}
Where $\mathcal{D}[L](\rho) = L\rho L^\dag - (L^\dag L \rho + \rho L^\dag L)/2$. We perform direct numerical simulations of this master equation using the python package qutip \cite{johansson_qutip:_2012,johansson_qutip_2013} for small $N$, which is presented in \fref{fig:outeff}. However, the dimension of the Hilbert space scales exponentially with $N$, thus we need an approximate method for large $N$; we apply the mean-field approximation.

The mean field approximation is applied by finding the equation of motion of the following expectation values: $\nu_{\rm em} = \langle c_{\rm em} \rangle$ and $\varsigma_{{\rm em},j}^k = \langle \sigma_{{\rm em},j}^{k} \rangle$, where $k={\rm x,y,z}$ which correspond to the appropriate Pauli matrices, then assuming all higher order expectation values can be factorized, e.g. $\langle c_{\rm em} \sigma_{{\rm em},j}^{\rm x} \rangle = \nu_{\rm em} \varsigma_{{\rm em},j}^{\rm x}$. Applying this approximation we get the following equations of motion:
\begin{align}
\dot{\varsigma}_{{\rm em},j}^{\rm x}(t) & = 2g \varsigma_{{\rm em},j}^{\rm z} \Re(\nu_{\rm em}) - \gamma \varsigma_{{\rm em},j}^{\rm x} /2, \\
\dot{\varsigma}_{{\rm em},j}^{\rm y}(t) & =  2g \varsigma_{{\rm em},j}^{\rm z} \Im(\nu_{\rm em}) - \gamma \varsigma_{{\rm em},j}^{\rm y} /2, \\
\dot{\varsigma}_{{\rm em},j}^{\rm z}(t) & =  -2 g ( \varsigma_{{\rm em},j}^{\rm x} \Re(\nu_{\rm em}) +  \varsigma_{{\rm em},j}^{\rm y} \Im(\nu_{\rm em})) \nn \\
&- \gamma (1 + \varsigma_{{\rm em},j}^{\rm z}), \\
\dot{\nu}_{\rm em}(t) & = \sum_{j=1}^N g (\varsigma_{{\rm em},j}^{\rm x} + i\varsigma_{{\rm em},j}^{\rm y})\nu_{\rm em}/2 - \kappa \nu_{\rm em}/2.
\end{align}
The equations of motion have an important symmetry: assuming the ions start in the same state, they will remain in the same state. Furthermore we are primarily interested in the average state of the ions, specifically: $\bar{\varsigma}_{{\rm em}}^{k} = \sum_{j=1}^N \varsigma_{{\rm em},j}^{k}/N$. Assuming that the initial state of the emitter is $|\psi_{\theta_0,\phi_0}\rangle$, we can simplify the equations of motion to:
\begin{align}
\dot{\bar{\varsigma}}_{{\rm em}}^{\rm x}(t) & = 2g \bar{\varsigma}_{{\rm em}}^{\rm z} \Re(\nu_{\rm em}) - \gamma \bar{\varsigma}_{{\rm em}}^{\rm x} /2, \label{eqn:meaneffs} \\
\dot{\bar{\varsigma}}_{{\rm em}}^{\rm y}(t) & =  2g \bar{\varsigma}_{{\rm em}}^{\rm z} \Im(\nu_{\rm em}) - \gamma \bar{\varsigma}_{{\rm em}}^{\rm y} /2, \\
\dot{\bar{\varsigma}}_{{\rm em}}^{\rm z}(t) & =  -2 g ( \bar{\varsigma}_{{\rm em}}^{\rm x} \Re(\nu_{\rm em}) +  \bar{\varsigma}_{{\rm em}}^{\rm y} \Im(\nu_{\rm em})) - \gamma (1 + \bar{\varsigma}_{{\rm em}}^{\rm z}), \\
\dot{\nu}_{\rm em}(t) & = N g (\bar{\varsigma}_{{\rm em}}^{\rm x} + i\bar{\varsigma}_{{\rm em}}^{\rm y})\nu_{\rm em}/2 - \kappa \nu_{\rm em}/2.  \label{eqn:meanefff}
\end{align}
We note that the mean-field makes a nonphysical prediction, with regard to the coupling between the cavity and the ions, that an unstable fixed point exists. In particular, if we set $\gamma=0$ and have an initial condition of $|\psi_{\theta_0,\phi_0}\rangle$ with $\theta_0 = 0$, which corresponds to $\bar{\varsigma}_{{\rm em}}^{\rm x} = \bar{\varsigma}_{{\rm em}}^{\rm y} = \nu_{\rm em} = 0$ and $\bar{\varsigma}_{{\rm em}}^{\rm z} = -1$, the mean-field predicts $\dot{\nu}_{\rm em} =  \dot{\bar{\varsigma}}_{{\rm em}}^{k} = 0$. This is nonphysical; if the ions start in an excited state the excitation will enter the cavity and be emitted. Physically, this process involves correlations forming between the ions and cavity, which the mean-field has neglected. This results in the mean-field being overly conservative in its prediction of efficiency and fidelity, making it best thought of as a lower-bound. The numerical solutions to \eref{eqn:meaneffs}-\eref{eqn:meanefff} presented in \fref{fig:outeff} were completed with the python package scipy \cite{walt_numpy_2011}.

\section{Simulations for fidelity of transfer} \label{app:fidtrans}

We use two methods to simulate the full state transfer and determine the fidelity: a direct truncation method and a mean-field method.

To perform a direct simulation for $N=1$, we simulate the non-Hermitian unnormalised wave equation of the total system \cite{gardiner_quantum_2004}:
\begin{align}
\frac{d|\tilde{\psi}(t)\rangle}{dt} = (-i H_{\rm total} - L_{\rm total}^\dag L_{\rm total}/2)|\tilde{\psi}\rangle. \label{eqn:nonhunneqn}
\end{align}
We are considering the $N=1$ case where all parts of the system are initially prepared in their respective ground states except the emitter. Consequently, there will be at most one excitation in the system at any given time. This allows us to truncate the wavefunction to the following form
\begin{align}
|\tilde{\psi}(t)\rangle = & \Big(\psi_{\mv{0}} + \int_{-\Xi}^{\Xi} d\xi \psi_{a}(\xi) a^\dag(\xi) \nn \\
& + \sum_{l={\rm em,re}} ( {\psi}_{\sigma,l} \sigma^\dag_{l}  + {\psi}_{c,l} c_{l}^\dag )  \Big) |\mv{0} \rangle.
\end{align}
Where $\psi_{\mv{0}}(t),\psi_{a}(\xi,t),{\psi}_{\sigma,l}(t),{\psi}_{c,l}(t)$ are wavefunction coefficients and $|\mv{0} \rangle$ refers to a state where the ions are in the ground state with the GEM and cavities in their vacuum state. The linear dynamical equations for these coefficients are:
\begin{align}
\dot{\psi}_{\sigma,{\rm em}}(t) = &  g \psi_{c,{\rm em}} - \gamma \psi_{\sigma,{\rm em}} / 2, \label{eqn:exs} \\
\dot{\psi}_{c,{\rm em}}(t) = & -g {\psi}_{\sigma,{\rm em}} - \kappa {\psi}_{c,{\rm em}} / 2, \\
\dot{\psi}_{a}(\xi,t) = & -i s(t) \xi \psi_{a}(\xi) - \zeta \int_{-\Xi}^\xi d\xi' {\psi}_{a}(\xi') \nn \\
& + i\sqrt{\zeta \kappa} {\psi}_{c,{\rm em}}, \label{eqn:exgem} \\
\dot{{\psi}}_{c,{\rm re}}(t) = &  g {\psi}_{\sigma,{\rm re}} - \kappa {\psi}_{c,{\rm re}}/2 + \sqrt{\zeta\kappa} {\psi}_{a}',\\
\dot{{\psi}}_{\sigma,{\rm re}}(t) = & -g {\psi}_{c,{\rm re}}(t) - \gamma {\psi}_{\sigma,{\rm re}} /2, \\
\dot{{\psi}}_{\mv{0}}(t) = & 0. \label{eqn:exf}
\end{align}
Where $\psi_{a}'(t) = S_{PP}(t) \int_{-\Xi}^{\Xi} d\xi \psi_{a}(\xi)$.

Normally the non-Hermitian wave equation must be stochastically simulated many times and averaged to get the density matrix of the system \cite{gardiner_quantum_2004}. However after a jump occurs the wave equation enters the state $|\mv{0} \rangle$, which is a dark steady state for the non-Hermitian Hamiltonian. In this special case, we only have to simulate \eref{eqn:nonhunneqn} once, then the density matrix for the total system is $\rho_{\rm total}(t) = P_{\rm jump} |\mv{0} \rangle \langle \mv{0} | + |\tilde{\psi} \rangle \langle \tilde{\psi}|$ where $P_{\rm jump} = 1 - \langle \tilde{\psi}| \tilde{\psi} \rangle$. We can use this density matrix to calculate the transfer fidelities.

For large $N$ we use a mean-field approximation to estimate the fidelity of the transfer. We start with the master equation for the total system
\begin{align}
\frac{d\rho_{\rm total}(t)}{dt} = & -i [H_{\rm total},\rho_{\rm total}] + \mathcal{D}[L_{\rm total}](\rho) \nn \\
& + \sum_{l={\rm em,re}} \sum_{j=1}^N \mathcal{D}[L^{\rm loss}_{l,j}](\rho). \label{eqn:masttot}
\end{align}
Using the same notation and technique as \eref{eqn:meaneffs}-\eref{eqn:meanefff} we get the following equations of motion:
\begin{align}
\dot{\bar{\varsigma}}_{{\rm em}}^{\rm x}(t) = & 2g \bar{\varsigma}_{{\rm em}}^{\rm z} \Re(\nu_{\rm em}) - \gamma \bar{\varsigma}_{{\rm em}}^{\rm x} /2, \label{eqn:mfs} \\
\dot{\bar{\varsigma}}_{{\rm em}}^{\rm y}(t) = &  2g \bar{\varsigma}_{{\rm em}}^{\rm z} \Im(\nu_{\rm em}) - \gamma \bar{\varsigma}_{{\rm em}}^{\rm y} /2, \\
\dot{\bar{\varsigma}}_{{\rm em}}^{\rm z}(t) = & -2 g ( \bar{\varsigma}_{{\rm em}}^{\rm x} \Re(\nu_{\rm em}) +  \bar{\varsigma}_{{\rm em}}^{\rm y} \Im(\nu_{\rm em})) \nn \\ 
& - \gamma (1 + \bar{\varsigma}_{{\rm em}}^{\rm z}), \\
\dot{\nu}_{\rm em}(t) = & N g (\bar{\varsigma}_{{\rm em}}^{\rm x} + i\bar{\varsigma}_{{\rm em}}^{\rm y})\nu_{\rm em}/2 - \kappa \nu_{\rm em}/2,  \\ 
\dot{\alpha}(\xi,t) = & -i s(t) \xi \alpha(\xi,t) - \zeta \int_{-\Xi}^\xi d\xi' \alpha(\xi',t) \nn \\
& + \sqrt{\zeta \kappa} \nu_{\rm em}, \label{eqn:mfgem} \\
\dot{\nu}_{\rm re}(t) = & N g (\bar{\varsigma}_{{\rm re}}^{\rm x} + i\bar{\varsigma}_{{\rm re}}^{\rm y})\nu_{\rm re}/2 - \kappa \nu_{\rm re}/2 + \sqrt{\zeta\kappa} \alpha',  \\ 
\dot{\bar{\varsigma}}_{{\rm re}}^{\rm x}(t) = & -2g \bar{\varsigma}_{{\rm re}}^{\rm z} \Re(\nu_{\rm re}) - \gamma \bar{\varsigma}_{{\rm re}}^{\rm x} /2, \\
\dot{\bar{\varsigma}}_{{\rm re}}^{\rm y}(t) = &  -2g \bar{\varsigma}_{{\rm re}}^{\rm z} \Im(\nu_{\rm re}) - \gamma \bar{\varsigma}_{{\rm re}}^{\rm y} /2, \\
\dot{\bar{\varsigma}}_{{\rm re}}^{\rm z}(t) = &  2 g ( \bar{\varsigma}_{{\rm re}}^{\rm x} \Re(\nu_{\rm re}) +  \bar{\varsigma}_{{\rm re}}^{\rm y} \Im(\nu_{\rm re})) - \gamma (1 + \bar{\varsigma}_{{\rm re}}^{\rm z}). \label{eqn:mff}
\end{align}
Where $\alpha' = S_{PP}(t) \int_{\Xi}^\Xi d\xi \alpha(\xi,t)$ and $\alpha(\xi,t) = \langle a(\xi)\rangle$.

We can calculate the transfer fidelity from the mean-field expectations by taking advantage of the Pauli operator density matrix factorization: $\bar{\rho} = (\sigma^x \bar{\varsigma}^x + \sigma^y \bar{\varsigma}^y + \sigma^z \bar{\varsigma}^z + I)/2$. The initial state will be pure, hence we can factorize it as $ \bar{\rho}_{\rm em}(0) = |\bar{\psi}_{\rm em} \rangle \langle \bar{\psi}_{\rm em}|$, which we can use to simplify the fidelity to $\mathcal{F} = \langle \bar{\psi}_{\rm em} (0)| \bar{\rho}_{\rm re}(2T) | \bar{\psi}_{\rm em} (0) \rangle$. Replacing the Pauli expansion for the emitter density matrix gives us the expression:
\begin{align}
\mathcal{F} = & (\bar{\varsigma}^x_{\rm em}(0) \bar{\varsigma}^x_{\rm re}(2T) + \bar{\varsigma}^y_{\rm em}(0) \bar{\varsigma}^y_{\rm re}(2T) \nn \\
& + \bar{\varsigma}^z_{\rm em}(0) \bar{\varsigma}^z_{\rm re}(2T) + 1)/2 \label{eqn:sigfid}
\end{align}

Simulations presented  of \eref{eqn:exs}-\eref{eqn:exf} and \eref{eqn:mfs}-\eref{eqn:mff} in \fref{fig:transevo} were performed with the differential equation package XMDS2 \cite{dennis_xmds2:_2013}. In order to improve numerical efficiency for the fidelity scans, simulations of \eref{eqn:exs}-\eref{eqn:exf} and \eref{eqn:mfs}-\eref{eqn:mff} presented in \fref{fig:transfid} were performed using the broadband solution of the GEM \eref{eqn:gemfsoln} instead of numerically solving \eref{eqn:exgem} and  \eref{eqn:mfgem}, and were performed using the python package scipy \cite{walt_numpy_2011}. 

\section{Simulations of state transfer with inhomogeneous broadening} \label{app:inhomotrans}

To simulate the state transfer with inhomogeneous broadening, we again make a mean field approximation on the master equation \eref{eqn:masttot}, except we use the modified Hamiltonian \eref{eqn:hamilinhomo}. This results in the following equations of motion:
\begin{align}
\dot{\varsigma}_{{\rm em},j}^{\rm x}(t) = & -2\Delta_{{\rm em},j} \varsigma_{{\rm em},j}^{\rm y}+2g \varsigma_{{\rm em},j}^{\rm z} \Re(\nu_{\rm em}) \nn \\
& - \gamma \varsigma_{{\rm em},j}^{\rm x} /2, \\
\dot{\varsigma}_{{\rm em}}^{\rm y}(t) = & 2\Delta_{{\rm em},j} \varsigma_{{\rm em},j}^{\rm x} + 2g \varsigma_{{\rm em},j}^{\rm z} \Im(\nu_{\rm em}) \nn \\
&- \gamma \varsigma_{{\rm em},j}^{\rm y} /2, \\
\dot{\varsigma}_{{\rm em},j}^{\rm z}(t) = &  -2 g ( \varsigma_{{\rm em},j}^{\rm x} \Re(\nu_{\rm em}) + \varsigma_{{\rm em},j}^{\rm y} \Im(\nu_{\rm em})) \nn \\
& - \gamma (1 + \varsigma_{{\rm em},j}^{\rm z}), \\
\dot{\nu}_{\rm em}(t) = & \sum_{j=1}^N g (\varsigma_{{\rm em},j}^{\rm x} + i\varsigma_{{\rm em},j}^{\rm y})\nu_{\rm em}/2 - \kappa \nu_{\rm em}/2,  \\ 
\dot{\alpha}(\xi,t) = & -i s(t) \xi \alpha(\xi,t) - \zeta \int_{-\Xi}^\xi d\xi' \alpha(\xi',t) \nn \\
& + \sqrt{\zeta \kappa} \nu_{\rm em}, \\
\dot{\nu}_{\rm re}(t) = & \sum_{j=1}^N g (\varsigma_{{\rm re},j}^{\rm x} + i\varsigma_{{\rm re},j}^{\rm y})\nu_{\rm re}/2 - \kappa \nu_{\rm re}/2 \nn \\
& + \sqrt{\zeta\kappa} \alpha',  \\ 
\dot{\varsigma}_{{\rm re},j}^{\rm x}(t) = & -2\Delta_{{\rm re},j} \varsigma_{{\rm em},j}^{\rm y} -2g \varsigma_{{\rm re},j}^{\rm z} \Re(\nu_{\rm re}) - \gamma \varsigma_{{\rm re},j}^{\rm x} /2, \\
\dot{\varsigma}_{{\rm re},j}^{\rm y}(t) = & 2\Delta_{{\rm re},j} \varsigma_{{\rm em},j}^{\rm x} -2g {\varsigma}_{{\rm re}}^{\rm z} \Im(\nu_{\rm re}) - \gamma \varsigma_{{\rm re},j}^{\rm y} /2, \\
\dot{\varsigma}_{{\rm re},j}^{\rm z}(t) = &  2 g ( \varsigma_{{\rm re},j}^{\rm x} \Re(\nu_{\rm re}) +  \varsigma_{{\rm re},j}^{\rm y} \Im(\nu_{\rm re})) \nn \\
& - \gamma (1 + \varsigma_{{\rm re},j}^{\rm z}). 
\end{align}
Where $\Delta_{l,j}$ are random variables sampled from the spectral density function $\rho_l(\Delta)$. In the limit of large $N$ we can take the continuum limit and change the equations to:
\begin{align}
\dot{\varsigma}_{{\rm em}}^{\rm x}(\Delta,t) = & -2\Delta \varsigma_{{\rm em}}^{\rm y}(\Delta)+2g \varsigma_{{\rm em}}^{\rm z}(\Delta) \Re(\nu_{\rm em}) \nn \\
& - \gamma \varsigma_{{\rm em}}^{\rm x}(\Delta) /2, \label{eqn:inhomoa} \\
\dot{\varsigma}_{{\rm em}}^{\rm y}(\Delta,t) = & 2\Delta \varsigma_{{\rm em}}^{\rm x}(\Delta) + 2g \varsigma_{{\rm em}}^{\rm z}(\Delta) \Im(\nu_{\rm em}) \nn \\ 
&- \gamma \varsigma_{{\rm em}}^{\rm y}(\Delta) /2, \\
\dot{\varsigma}_{{\rm em}}^{\rm z}(\Delta,t) = & -2 g ( \varsigma_{{\rm em}}^{\rm x}(\Delta) \Re(\nu_{\rm em}) + \varsigma_{{\rm em}}^{\rm y}(\Delta) \Im(\nu_{\rm em})) \nn \\
& - \gamma (1 + \varsigma_{{\rm em}}^{\rm z}(\Delta)), \\
\dot{\nu}_{\rm em}(t) = & \int_{-\infty}^{\infty} d\Delta \; \varrho_{\rm em}(\Delta) g (\varsigma_{{\rm em}}^{\rm x}(\Delta) + i{\varsigma}_{{\rm em}}^{\rm y}(\Delta))\nu_{\rm em}/2 \nn \\
&- \kappa \nu_{\rm em}/2,  \\ 
\dot{\alpha}(\xi,t) = & -i s(t) \xi \alpha(\xi,t) - \zeta \int_{-\Xi}^\xi d\xi' \alpha(\xi',t) \nn \\
& + \sqrt{\zeta \kappa} \nu_{\rm em}, \\
\dot{\nu}_{\rm re}(t) = & \int_{-\infty}^{\infty} d\Delta \;\varrho_{\rm re}(\Delta) g (\varsigma_{{\rm re}}^{\rm x}(\Delta) + i\varsigma_{{\rm re}}^{\rm y}(\Delta))\nu_{\rm re}/2 \nn \\
& - \kappa \nu_{\rm re}/2 + \sqrt{\zeta\kappa} \alpha'  \\ 
\dot{\varsigma}_{{\rm re}}^{\rm x}(\Delta,t) = & -2\Delta \varsigma_{{\rm re}}^{\rm y}(\Delta) -2g \varsigma_{{\rm re}}^{\rm z}(\Delta) \Re(\nu_{\rm re}) \nn \\
&- \gamma \varsigma_{{\rm re}}^{\rm x}(\Delta) /2, \\
\dot{\varsigma}_{{\rm re}}^{\rm y}(\Delta,t) = & 2\Delta \varsigma_{{\rm re},j}^{\rm x} -2g  {\varsigma}_{{\rm re}}^{\rm z}(\Delta) \Im(\nu_{\rm re}) \nn \\
&- \gamma \varsigma_{{\rm re}}^{\rm y}(\Delta) /2, \\
\dot{\varsigma}_{{\rm re}}^{\rm z}(\Delta,t) = & 2 g ( \varsigma_{{\rm re},j}^{\rm x}(\Delta) \Re(\nu_{\rm re}) +  \varsigma_{{\rm re}}^{\rm y}(\Delta) \Im(\nu_{\rm re})) \nn \\
&- \gamma (1 + \varsigma_{{\rm re}}^{\rm z}(\Delta)), \label{eqn:inhomoz}
\end{align}
where $\varrho_l(\Delta)$ is normalized to $\int_{-\infty}^{\infty}d\Delta \; \varrho_{l}(\Delta) = N$ and the ion variables $\varsigma_l^k(\Delta,t)$ are now a function of $\Delta$ and $t$. In the continuum limit, the excitation spectral density is $E_{\sigma, l}(\Delta,t) = \varrho_{l}(\Delta)(\varsigma_{l}^{\rm z}(\Delta,t)+1)/2$ and the average operators are 
\begin{align}
\bar{\varsigma}_{l}^{k}(t) = & \frac{1}{N} \int_{-\infty}^{\infty} d\Delta \; \varrho_{l}(\Delta) \varsigma_l^k(\Delta,t).
\end{align}
Fidelities can then be calculated with \eref{eqn:sigfid}. Simulations of \eref{eqn:inhomoa}-\eref{eqn:inhomoz} were performed with the python package scipy \cite{walt_numpy_2011} using the broadband solution of GEM \eref{eqn:gemfsoln} to improve numerical efficiency, the results are presented in \fref{fig:finhomo} and \fref{fig:finhomoscan}. 

\bibliographystyle{apsrev4-1}
\bibliography{Zotero}

\end{document}